\documentstyle[11pt,epsfig]{article}

\begin{document}

\begin{center}

{\Large Comment to the paper  D. V. Anchishkin and S. N. Yezhov  \\
``Thermalization in Heavy-ion Collisions''}%

\vspace{1.0cm}

{\bf K. A. Bugaev}%

\vspace{1.0cm}

{Bogolyubov Institute for Theoretical Physics of the Academy of Science of Ukraine}%

e-mail: bugaev@th.physik.uni-frankfurt.de

\date{\today}

\end{center}

\vspace{1.0cm}

\begin{abstract}
Here I discuss  the major  pitfalls and the most severe  mistakes of the above mentioned paper. 
The thorough analysis shows that despite all the claims the present work 
has nothing to do with the thermalization process in relativistic heavy ion collisions. 
In contrast to the authors' beliefs I show that their main result is not derived, but is  a combination of mathematical mistakes and hand waving arguments. 
\end{abstract}

\vspace{4.cm}

\noindent
{PACS: 25.75.-q, 25.75.Nq}\\
{\bf Key words:} Thermalization, heavy ion physics, nonequilibrium distribution functions

\newpage

Thermalization phenomenon observed in heavy ion collisions at intermediate and
high energies is one of the pillars of modern nuclear physics.  
It means  a  transformation  of two colliding pieces of  cold nuclear matter,  
which produce
the  highly excited
non-equilibrated strongly interacting matter, into  a media being in local thermal  equilibrium. 
There exist 
a large number of theoretical  works devoted to its explanation  on a basis of various approaches and models. The latter are varying from the gluon transport model \cite{Thermaliz:1} and
hadronic cascade model \cite{Thermaliz:2,Thermaliz:3} to rather sophisticated hydro-kinetic approaches
\cite{Bugaev:02}.

In the present work \cite{anchish:08} the authors claimed no less than  they found ``an explanation for the thermalization 
phenomenon, starting from such basic concepts as the conservation laws of energy and momentum and leaving the details which are characteristic of every specific process, aside for further researches''. However, their model, {\it the maximal isotropization model (MIM)}, has nothing to do with the collisions of heavy ions and, as shown below,
does not shed any light on a thermalization process at all. 

The authors, indeed, account for the energy and momentum conservation in all two particle collisions and in this way they redistribute the total energy and momentum of the system among the ideal  Boltzmann particles. However, the authors have neglected any inelastic reaction in the system and production of any particle in it. Therefore, from the beginning the possible range of applications, if we consider the hadrons,  is restricted 
from above by the { lab.   energy  per particle which must be below the pion mass 
$m_\pi \approx 140 $ MeV.  However, the authors of  \cite{anchish:08} continue to use the relativistic expression
for particle energy and even apply their model to pions with the mean energy up to $800 $ MeV.
After I realized this, it became clear to me that the thorough  discussion of such `tiny issues' as 
spherical symmetry  of the momentum distribution after each elastic collision assumed by authors, 
or an absence of any cross-sections in all their equations, or the energy and time  dependence of the 
momentum integration volume $V_p$, introduced by them, e.t.c., would require too much time,  and hence 
I decided to analyze just a few most important issues.
} 

{\bf \small  
}

The authors operate with the simple equations which are well known to the community
\cite{Aichelin, GorenKostyuk}, 
but fail to analyze  or even simply  to  discuss  the validity of their main 
approximation made in Eq. (15).  However, the approximation suggested by the authors 
drastically changes the properties of the system from the normal ones to the anomalous
ones. It is well known from the standard course of statistical mechanics (see, for instance, paragraph 67 of \cite{Rumer})  that, if the microcanonical density of states increases or decreases as a certain  power of particle energy, then the system has the normal properties 
since the corresponding canonical  system can have positive temperatures only.
Unfortunately,  the microcanonical density of  states  obtained  after the approximation (15) for fixed values of $a_n$ and $b_n$ is exponentially decreasing function of particle  energy and hence it   leads to the anomalous system properties since it allows 
the existence of negative temperatures \cite{Rumer}. This mistake has the  fatal consequences for the present paper.

However, the wrong consequences of the approximation (15) are  not seen immediately because the authors make another mistake by
 loosing  their original object of analysis. 
Thus,  originally the authors are dealing with the 
microcanonical distribution (11) and then suddenly they make the Laplace transform 
to the canonical density of states in Eq. (12). The rest of the paper is devoted to the analysis of the canonical distribution (12) with the temperature of the external thermostat 
given by the parameter of the Laplace transform $ T = \beta^{-1}$ which afterwards   appears in all relevant expressions. Thus, 
instead of elucidating the temperature of the microcanonical system which would be necessary for studying the thermalization phenomenon,  the authors insert 
this very temperature  $ T = \beta^{-1}$
by hands through the canonical partition. Therefore,  {\bf the present  work does not give any  novel  piece of  correct information on the thermalization process  in heavy ion collisions.}

Perhaps, the authors of MIM do not know that, when the Laplace transform (12) is made,
it is assumed that there exists an ensemble of microcanonical systems 
of all  allowed energies and each of  these systems is  brought into a contact with the external thermostat whose  temperature is  $ T = \beta^{-1}$ 
(for a detailed discussion see  \cite{HagedornTH:1,HagedornTH:2}). It seems that the authors of this work  do not know that the microcanonical 
ensemble allows one to define the microcanonical temperature $T_{mce}$ 
via  the derivative of the entropy $S_{mce}$ with respect to the total energy $E_{tot}$  
in a standard way as  $T_{mce}^{-1} = \left( \frac{\partial S_{mce} }{\partial E_{tot} } \right)_{V,N}$ \cite{HagedornTH:1,HagedornTH:2}. If they applied  such a formula  to the 
microcanonical density of states, it would not be necessary to write the wrong 
and meaningless discussion given in  Sect. 3.4  (see below).

Moreover, the authors do not understand that to elucidate the temperature of 
the system of colliding particles one has to analyze the microcanonical density of 
states, since in the canonical ensemble the outer thermostat forces system to 
adopt its temperature. The Laplace transform 
 to the canonical ensemble is 
an auxiliary  mathematical trick to simplify the calculation of the original microcanonical partition  with the  constraints (the conservation laws in this work). After the constraints are evaluated in the canonical ensemble, 
one has to perform the inverse Laplace   transform 
to the original ensemble
and analyze it. Some general results on the both Laplace transforms  
in the semiinfinite  or finite  interval can be found in  Refs. \cite{Laplace:1}
and \cite{Laplace:2}, respectively.  
I am sure  that these references would be very useful to the authors of \cite{anchish:08},
since they do not know that in the right hand side of the inverse Laplace transform
there must be an additional factor $\frac{1}{2\pi i}$, which is missing
in their Eqs. (52), (53) and in the discussion presented between Eqs.  (51) and (53) (see Sect. 3.4.). 
{This mistake  was extremely  surprising to me since  one of the authors of Ref. \cite{anchish:08} 
was present  at  many of  seminars  of mine where the results of works  \cite{Laplace:1} and
\cite{Laplace:2} just devoted to  various applications of  the Laplace transform technique were discussed in great details!

Furthermore, the authors of  \cite{anchish:08}  claim that they  {\bf derived}  the nonequilibrium momentum distribution function. The same statement is repeated in their subsequent work \cite{anchish:08b}.
As was shown above none of the distributions listed in Ref.  \cite{anchish:08}, was actually derived. 
Moreover, by claiming this the authors clearly indicate an absence of any knowledge of great amount 
of works devoted to studies of different aspects of  nonequilibration  phenomena occurring  in relativistic 
high energy nuclear collisions. The authors of Ref.   \cite{anchish:08} implicitly mean that the `derived'  nonequilibrium momentum distribution function has some correction to the  Boltzmann momentum distribution 
of relativistic particles of ideal gas.  Note, however, that more realistic interaction between relativistic 
particles (reltivistic hard-core repulsion)  leads to a far more complicated momentum distribution 
\cite{RelVDW} than the Boltzmann one and this fact  has nothing to do with 
the collision of two relativistic nuclei, but  is an inherent property of the equation of state of hadronic matter. Thus,  accounting for realistic interaction between hadrons one obtains
the modification of the Boltzmann momentum distribution. 
It is also  well known that the non-Boltzmann momentum distribution of the secondary hadrons formed in the high energy  nuclear  collisions can be a consequences of a complicated  collective motion of particles 
\cite{Collect} which  have 
the Boltzmann distribution in their local rest frame,  but are measured by detector in a lab. frame.
In addition, the conditions of a kinetic  freeze-out process of the secondaries also affect  their  momentum spectra \cite{FO:1,FO:2,FO:3, FO:4,Bugaev:02} and this modifies the Boltzmann distribution of particles in the fluid  elements 
 freezing  out  at the time-like  freeze-out hypersurfaces \cite{FO:2,FO:3}. 
}

From the discussion above 
it is evident, if the authors of  \cite{anchish:08}  applied  the standard thermodynamic relation to the microcanonical density of states obtained  from the canonical one  after the inverse Laplace transform over $\beta$ with 
$\exp[  \beta E_{tot}]$, then they 
would 
get  the result independent  of the  Laplace transformation parameter $\beta =  T^{-1}$. 
Hence I conclude that   the main result of the present paper does not correspond to the collisions
of heavy ions at all, but it represents the mixing of two (directed to each other) slow flows of small hard balls in a box.
The box and balls are in a thermal contact with an external thermostat while  the interaction 
with the box's  walls is neglected. 
Thus, the present work does not corresponds to its title and to the claims written
in its introduction. Moreover, it scientific value seems to be negative. 

As explained  above,  the MIM does not account for particle production and, hence,
it cannot be used to relativistic nuclear  collisions. 
Nevertheless, without any explaining remark the authors apply their analysis 
to the ideal (Boltzmann)  gas of pions at ultrarelativistic energies per particle  (up to 800 MeV). 
However, besides other mistakes,  this implies that the heavy ions consist  of free pions.

All this is not surprising because, in contrast to  
the MIM,  in high energy  nuclear collisions a  thermalization occurs not only due to elastic  scattering of the particles, but also and mainly due to production of new (sorts of) particles \cite{Thermalization,Thermaliz:3} which automatically increases the 
local particle density in a system which, in its turn, essentially enhances the reaction rate
leading to a thermalization. 
Since such a production of particles  is not and cannot be accounted within the framework of  the present work, its 
output is unphysical.

Also I have to stress a very sloppy style of presentation and its mentor tone which 
are in a tremendous  dissonance with the trivial mistakes and pitfalls of this work. 
Thus,  one hand the present authors  pretend on a solving  a fundamental problem,   but on the other hand  they are  not careful with their definitions, approximations and even wording. After I read a few paragraphs and did not find any word of what kind of statistics is used, I was not already  surprised by such pearls as ``the confinement of 
the momentum space'' or ``the nonnormalized distribution"  (see the sentence after Eq. (10)) and so on.

In summary, the most original part of  present work is a collection of primitive mistakes
which have  nothing to do with the  thermalization problem in heavy ion collisions.


\begin{thebibliography}{99}  
  
\bibitem{Thermaliz:1}  
%
More references can be found in 
A. H. Mueller, A. I. Shoshi and S. M. H. Wong,
Phys. Lett. {\bf B 632}, 257 (2006).  

  
\bibitem{Thermaliz:2}
%
%
see, for instance, 
M. Belkacem et al.,  
Phys. Rev. {\bf C 58},  1727
(1998).

\bibitem{Thermaliz:3}
%
L. Bravina  et al., 
Int. J. Mod. Phys. {\bf E 16}, 777 (2007).

\bibitem{Bugaev:02}
K. A. Bugaev, 
Phys. Rev. Lett. {\bf 90},  252301 (2003);
Phys. Rev. {\bf C 70},  034903  (2004) and references therein.



 
\bibitem{anchish:08}
%
D. V. Anchishkin and S. N. Yezhov, 
Ukr.  J. Phys. {\bf 53},  87 (2008).
  

\bibitem{Aichelin}
%
for more specialized references see, for instance, 
K. Werner and J. Aichelin, 
Phys.Rev. {\bf C 52}, 1584 (1995);
F. M. Liu, K. Werner  and J. Aichelin,
Phys. Rev. {\bf C 68}, 024905  (2003). 
 




\bibitem{GorenKostyuk}
%
see also V.V. Begun, M. I. Gorenstein,  A. P. Kostyuk, and O. S. Zozulya,
Phys. Rev. {\bf C 71}, 054904 (2005) and references therein.
 


\bibitem{Rumer}
Yu. B. Rumer and M. Sh. Ryvkin, ``Thermodynamics, Statistical Physics and Kinetics'',
Second Edition (Russian), Nauka, Moscow, 1977.

\bibitem{HagedornTH:1}
%
L. G. Moretto,  K. A. Bugaev, J. B. Elliott and L. Phair,
{Europhys. Lett. } {\bf 76},  402 (2006);
%
K. A. Bugaev,  J. B. Elliott, L. G. Moretto and  L. Phair,
hep-ph/0504011.


\bibitem{HagedornTH:2}
%
L.~G.~Moretto, L.~Phair, K.~A.~Bugaev and J.~B.~Elliott,
{\bf  PoS   CPOD2006:037}, (2006)  18p.
%
L.~G.~Moretto, K.~A.~Bugaev, J.~B.~Elliott and L.~Phair,
nucl-th/0601010. 

\bibitem{Laplace:1}
%
K. A. Bugaev et. al., 
Phys. Rev. {\bf C 62},  044320 (2000);
nucl-th/0007062;
Phys. Lett. {\bf B 498}, 144  (2001);
nucl-th/0103075;
P. T. Reuter and K. A. Bugaev,
Phys. Lett. {\bf B 517},  233 (2001);
{Ukr. J. Phys.}, {\bf 52},  489  (2007);
K. A. Bugaev, 
{Phys. Rev.} {\bf C 76},  014903   (2007);
arXiv:0711.3169 [hep-ph]; 
K. A. Bugaev, V. K. Petrov and G. M. Zinovjev,
{arXiv:0801.4869} [hep-ph] and
{arXiv:0807.2391} [hep-ph].

 


\bibitem{Laplace:2}
%
K. A. Bugaev,
{ Acta. Phys. Polon.} {\bf B 36},  3083 (2005);
nucl-th/0507028; 
K. A. Bugaev, L. Phair and J. B. Elliott,
{ Phys. Rev.} {\bf E 72},  047106  (2005); 
K. A. Bugaev, 
Phys. Part. Nucl. {\bf 38},  447 (2007).


\bibitem{anchish:08b}
D. V. Anchishkin and S. N. Yezhov, 
arXiv:0804.1745 [nucl-th]. 

\bibitem{RelVDW}
%
some examples can be found in
K. A. Bugaev, M. I. Gorenstein, H. St\"ocker and W. Greiner,
Phys. Lett. {\bf B 485}, 121  (2000);
G. Zeeb, K. A. Bugaev, P. T. Reuter and H. St\"ocker,
{Ukr.  J. Phys.}  {\bf 53},   279 (2008);
K. A. Bugaev, 
Nucl. Phys. {\bf A 807},   251 (2008).

\bibitem{Collect}
%
the full list of references can be found in 
K. A. Bugaev, M. Gazdzicki and M. I. Gorenstein,
Phys. Lett. {\bf B 523},  255  (2001);
Phys.  Lett. {\bf B 544},  127 (2002);
Phys. Rev. {\bf C 68},  017901 (2003); 
K. A. Bugaev,
J. Phys. {\bf G 28},  1981 (2002);
M. I. Gorenstein, K. A. Bugaev and  M. Gazdzicki,
Phys. Rev. Lett. {\bf 88},  132301 (2002).


\bibitem{FO:1}
%
M. I. Gorenstein and Yu. M. Sinyukov, Phys. Lett. {\bf B 142},  425 (1984);
{ Sov. Nucl. Phys. {\bf 41}}, 797  (1985) .


\bibitem{FO:2}
%
Yu. M. Sinyukov, {Sov. Nucl. Phys.} {\bf 50},   228 (1989);
Yu. M. Sinyukov, Z. Phys. {\bf C 43}, 401 (1989);
Yu. M. Sinyukov, S. V. Akkelin, Y. Hama, 
Phys. Rev.  Lett. {\bf 89},   052301 (2002).
M. S. Borysova  {\it et al.,}
 Phys. Rev. {\bf C 73},  024903 (2006).

\bibitem{FO:3}
%
K. A. Bugaev, 
Nucl. Phys. {\bf A 606},  559 (1996);
K. A. Bugaev, M. I. Gorenstein and W. Greiner,
J. Phys. {\bf G 25},  2147 (1999); 
Heavy Ion Phys. {\bf 10},  333  (1999);
K. A. Bugaev and M. I. Gorenstein, 
nucl-th/9903072. 

\bibitem{FO:4}
%
V. K. Magas et. al., 
Eur. Phys. J. {\bf C 30}, 255 (2003);
Nucl. Phys. {\bf A 749},  202 (2005).

\bibitem{Thermalization}
%
A. Dumitru  and C. Spieles,
Phys. Lett. {\bf B 446},  326 (1999) and references therein. 



\end{thebibliography}
\end{document}